\documentclass[twocolumn,11pt]{IEEEtran}
\usepackage{graphicx} % Required for inserting images
\usepackage[table]{xcolor}
\usepackage{mathrsfs}
\usepackage{hyperref}
\usepackage{cite}
\usepackage{authblk}
\usepackage{amsmath,amssymb,amsfonts,amsthm,bm,bbm,dsfont}
\usepackage{overpic}
\newcommand{\rcomment}[1]{\textcolor{red}{#1}}
\DeclareMathOperator{\der}{\Longrightarrow}
\DeclareMathOperator{\dedup}{\longrightarrow}
\DeclareMathOperator{\ccap}{\mathsf{cap}}

\renewcommand{\leq}{\leqslant}

\newcommand{\Z}{\mathbb{Z}}
\newcommand{\sA}{\mathsf{A}}
\newcommand{\sC}{\mathsf{C}}
\newcommand{\sG}{\mathsf{G}}
\newcommand{\sT}{\mathsf{T}}

\usepackage{mathtools}
\usepackage{xparse}
\DeclarePairedDelimiterX{\set}[1]{\{}{\}}{\setargs{#1}}
\DeclarePairedDelimiterX{\bset}[1]{[}{]}{\setargs{#1}}
\NewDocumentCommand{\setargs}{>{\SplitArgument{1}{;}}m}
{\setargsaux#1}
\NewDocumentCommand{\setargsaux}{mm}
{\IfNoValueTF{#2}{#1} {#1\,\delimsize|\,\mathopen{}#2}}%{#1\:;\:#2}

\DeclarePairedDelimiter\abs{\lvert}{\rvert}

\newtheorem{example}{Example}
\title{Error Correction for DNA Storage}
\date{March 2023}
\author{Jin~Sima,~%~\IEEEmembership{Member,~IEEE,}
	Netanel~Raviv,~\IEEEmembership{Senior Member,~IEEE,}

 Moshe Schwartz,~\IEEEmembership{Senior Member,~IEEE,}
 and~Jehoshua~Bruck,~\IEEEmembership{Life Fellow,~IEEE}% <-this % stops a space
	
	\thanks{J. Sima is with the Department of Electrical and Computer Engineering, University of Illinois Urbana-Champaign, Urbana 61801, IL, USA (e-mail: jsima@illinois.edu).}% <-this % stops a space
 
	\thanks{N. Raviv is with the Department of Computer Science and Engineering, McKelvey School of Engineering, Washington University in Saint Louis, St. Louis, MO, 63130 USA (email:  netanel.raviv@wustl.edu).}% <-this % stops a space
\thanks{M. Schwartz is with the Department of Electrical and Computer Engineering, 
McMaster University, Hamilton, ON, Canada (e-mail: schwartz.moshe@mcmaster.ca)}
\thanks{J. Bruck is with the Electrical Engineering Department, California Institute of Technology, Pasadena,
CA, 91125 USA (e-mail: bruck@caltech.edu)}
}

\begin{document}

\maketitle

\section{Introduction}\label{sec:introduction}

% \begin{itemize}
%     \item Data storage in general.
%     \item Information theory represents information using binary.
%     \item The notion of redundancy and why it should be optimized.
%     \item Magnetic storage 1956 (reference?) (picture?).
%     \item Challenges in current storage technology. Archival storage, ``cold'', data storage regulations.
%     \item Introduction of a DNA storage system.
%     \item ``Humans will not forget how to read DNA.''
%     \item Structure of the paper in natural language (duplication for in vivo, the others for in vitro). ``Evolution correcting codes.''
%     \item Keep current section order.
% \end{itemize}

Information theory is considered by many as the mathematical theory of \textit{communication}. Normally, the word ``communication'' describes a scenario involving two physically distant parties that exchange information, but may equally involve two \textit{temporally} distant parties that do so. The latter gives rise to communication across time, rather than across space, and is commonly referred to as \textit{information storage} \cite{berlekamp1980technology}, 
i.e., the process of encoding information into a physical device in order to retrieve it at a later point in time, efficiently and accurately. 

In his groundbreaking 1948 paper, Claude Shannon (1916-2001) showed that \textit{all} types of information (images, text, videos, etc.) can be communicated using bits, i.e., zeros and ones, and an identical statement holds in the case of storage. In order to store a piece of information, one has to encode it using bits, and place those bits on a reliable physical device, preferably a \textit{non-volatile} one, i.e., that does not require electric current to retain that information.

The earliest example of high-density non-volatile storage device (beyond punch-cards and written media which existed for millennia) is probably that of \textit{magnetic storage}. In this 1950's technology, bits were organized on a magnetizable tape using different magnetization patterns. Over the following decades, increased demand for higher storage volumes pushed this technology forward to become the \textit{hard-disk drives}, which in recent years cleared the way to \textit{solid-state drives}. 

Albeit over 10 orders of magnitude increase in volume since they were introduced, digital storage devices struggle to keep up with increasing storage demands. The immense volume of data generated today, especially since the emergence of information sharing platforms such as YouTube and social networks, is projected to pass the rate in which digital storage devices improve. Especially prominent is the growing requirement for ``cold'' storage, i.e., one which is seldom accessed, such as old family photos or historical records. One of the most promising and most radical new technologies to resolve the cold storage problem, is \textit{DNA storage}, i.e., storing information in DNA molecules.

In a way, DNA molecules are the storage device of nature. These long molecules, which contain sequences of either one of four basic molecules called nucleotides $\sA,\sC,\sG,\sT$, are used by all living organisms to communicate across time. The DNA molecules contain ``recipes'' for producing proteins, which are the building blocks of living organisms. By communicating DNA across generations, these recipes are literally transmitted from parents to offsprings, and enable life to continue.

In the past several decades, scientists have had remarkable success in creating artificial DNA molecules in the lab, and in keeping those molecules stable in a vial (i.e., not inside any living cell). The $\sA,\sC,\sG,\sT$ nucleotides in these artificial molecules can be chosen freely, and can therefore store bits just like any other storage device. For instance, one can decide that
\begin{align*}
    \sA=00, \sC=01, \sG=10, \sT=11
\end{align*}
and then store the sequence~$00101001$ by creating the DNA molecule~$\sA \sG \sG \sC$ in the lab. DNA storage has fantastic and far-reaching advantages over existing technology:
\begin{itemize}
    \item \textit{DNA storage is ultra-dense}. Current data centers are the size of buildings; storing similar amounts of data in DNA would require the size of a refrigerator.
    \item \textit{DNA storage is stable}. DNA molecules can last tens of thousands of years without any energy investment; some off-the-shelf hard drives will not be usable in as little as 20 years.
    \item \textit{DNA storage is future-proof}. As long as there are humans, DNA reading technology will be of interest, and DNA reading devices will exist. This can hardly be said about, say, floppy disks, whose reading nowadays requires a trip to a museum. In other words, while humans already almost forgot how to read some fairly recent storage devices, \textbf{humans will never forget how to read DNA}.
\end{itemize}
A typical process to store information in DNA is as follows: First, the data, represented by 0's and 1's, are encoded into sequences of nucleotides. Then, the DNA molecules containing these sequences of nucleotides are synthesized and stored either in vials or in living organisms, which is the data writing process. To read the data, the Polymerase Chain Reaction (PCR) technology is used to access the part of the data to be retrieved. Then, the DNA molecules obtained after the PCR process are read using DNA sequencing techniques, thereby recovering the sequences of nucleotides. Finally, the sequences of nucleotides are decoded back to the data. A more detailed description of a DNA storage workflow is given in Section \ref{section:sliced}.

Whether old or new, all storage devices are prone to \textit{errors}. Due to imperfect hardware, physical damage, or deterioration of materials, some bits in any storage device might be read in error. Without proper preparation, losing even a single bit might render the respective piece of information unreadable, and therefore lost. Coding-theorists and engineers have been combating this phenomenon ever since storage devices were invented, and various \textit{error-correction} mechanisms, known as \textit{codes}, were developed. 

All these mechanisms require adding \textit{redundancy} to the data, i.e., to store more bits than the actual size of the data. These redundant bits are then used in the reading process in cases where some bits are read in error. The simplest form of redundancy is \textit{replication}: instead of storing every bit of the data once, store it 3 times. For example,
\begin{align*}
    &\text{Data: } &&01001\\
    &\text{Store: } &&000111000000111
\end{align*}
Then, while reading a possibly error-filled sequence from the device, the most frequent bit among every consecutive triple is most likely the correct one:
\begin{align*}
    &\text{Store: } &&000111000000111\\
    &\text{Errors: } &&\rcomment{1}0011\rcomment{0}00000\rcomment{1}111\\
    &\text{Correction: } &&0\phantom{00}1\phantom{11}0\phantom{00}0\phantom{00}1\phantom{11}
\end{align*}

But how much redundancy is enough? The example above shows a three-times increase in the amount of storage, a high price to pay. The minimal amount of redundancy required to correct errors is an ongoing and difficult research area in coding theory, and depends on the type of storage medium. Coding theorists have worked relentlessly over the past 80 years in order to come up with algorithms that guarantee error-free information storage for the existing storage technology. However, as we shall see throughout this paper, DNA storage devices have a very unique structure and constraints, which give rise to new and interesting \textbf{types of errors which have never been studied}.

The error shown above is called \textit{substitution}: a ``1'' bit is replaced by a ``0'' bit, or vice versa. This is a common and well-understood error in traditional storage media, which seldom appears in DNA storage devices. However, most errors in DNA storage devices are new, i.e., have never appeared in traditional devices. These new types of errors depend on the type of DNA storage device at hand. 

DNA storage devices are partitioned to two families: the most common family is called \textit{in-vitro}, which includes devices that contain a vial with short, unordered sequences of DNA that float in a solution inside that vial. The other less common family is called \textit{in-vivo}, where artificially synthesized DNA molecules are planted inside a primitive life form, such as bacteria, for better data longevity and stability. Better longevity is guaranteed by the self-sustaining property of primitive life-forms; with minimal energy investment a bacteria colony could last millions of years. Better stability is guaranteed by reproduction across generations; redundancy in the data will be introduced naturally, and will be stabilized via natural selection.

%in a living organism it will live forever]  [natural redundancy via organism replicating and resistence via natural selection].

In-vitro DNA storage devices pose several interesting coding-theoretic challenges. First, they are prone to known but understudied errors called \textit{deletions}. In a deletion, a bit completely disappears from a sequence without leaving a trace, and the read sequence is shorter than the one which was initially stored. Deletions occur in DNA storage mainly due to imperfections in the synthesis reaction, which commonly skip one or more of the nucleotides that need to be written due to weak chemical bonds. A closely related type of errors, where redundant bits sneak into the sequence unexpectedly, are called \textit{insertions}.

Even though mechanisms for deletion correction, known as deletion codes, were studied to some extent since the 1960's, much was left unknown. Driven by the advent in DNA storage, the interest in deletion errors increased recently, and optimal solutions were found only in the past few years. Deletion errors will be described in Section~\ref{sec:delcodes}.

An even more substantial challenge in in-vitro DNA storage systems is the fact that only \emph{short and unordered} DNA sequences can be stored together in a vial. Current limitation in DNA-synthesis technology can only generate sequences that are a few thousand nucleotides long, and placing those sequences together in a vial makes it impossible to know which one comes after or before any other. This is in sharp contrast to traditional storage media, where data is partitioned to \textit{pages}, that always appear in memory in the same order they were written. 

A simple solution to the ordering problem comes in the form of \textit{indexing}: begin each short DNA sequence with several nucleotides that determine its correct position relative to other sequences in the same vial. Surprisingly, this is \textit{not} the best solution in terms of the amount of redundancy. Due to errors, the indexing nucleotides might get scrambled and interfere with the correct order. An optimal solution for the order problem was also found very recently, and it is described in Section~\ref{section:sliced}.

For in-vivo storage, however, the picture is remarkably different. While placing the synthetic DNA inside bacteria improves its longevity and stability, it exposes the stored data to the natural bio-chemical and evolutionary processes inside the bacteria. As the reader might already know, cells reproduce by a process called \textit{mitosis}, where one cell splits into two. During mitosis, DNA molecules replicate themselves in each one of the offspring cells, a process that is not  perfect, and some errors might occur. In nature, these errors are the basis of Darwin's natural selection theory: arbitrary errors cause arbitrary mutations, and only the mutations which improve the organism's ability to survive persist among generations. In the context of information storage, however, these errors must be understood, and corrected; an \emph{evolution-correcting code} must be developed. A common error in this setting is \textit{duplication}, where a piece of DNA material is replicated, and attaches itself at a different location. Duplication errors in in-vivo DNA storage are described in Section~\ref{section:duplication}.

%\newpage
\section{Deletion Codes}\label{sec:delcodes}
Though deletion errors were studied from the 1960's, motivated by synchronization errors in traditional media, the interest in correcting deletion errors increased recently due to their prevalence in DNA storage. 
As mentioned in Section \ref{sec:introduction}, deletions, insertions, and substitutions, are the notable three types of errors that occur in the reading, writing, and storing processes in DNA storage. 
%Moreover, deletions, insertions, substitutions are the three types of DNA mutations. 
Hence, codes correcting these three types of errors are necessary for reliably storing information in DNA. Beyond the applications in DNA storage and communication, the study of deletions, insertions, and substitutions, is also connected to edit distance and sequence alignment, etc., which have applications in natural language processing and other applications involving DNA sequence analysis.

\subsection{What are deletion, insertion, and substitution errors?}

We now describe the three types of errors in greater detail. A deletion removes a symbol from a sequence; in the context of DNA storage, it removes a nucleotide from the sequence, e.g., turning $\sT \sG \sG \sA$ into $\sT \sG \sA$. In the context of natural language, it removes a letter from a word or a sentence, e.g., turning the word ``cat'' into ``at.'' An insertion adds an extra symbol to the sequence, e.g., turning $\sA \sC \sT \sG$ into $\sA \sC \sC \sT \sG$ or turning ``eat'' into ``heat.'' A substitution replaces a symbol in the sequence, e.g., changing $\sT \sG \sG \sA$ to $\sT \sG \sG \sG$ or ``for'' to ``far.''
 
Among those three types of errors, substitution errors are better understood compared to deletions and insertions, as there are many classic code constructions for correcting substitution errors such as Hamming codes, polar codes, LDPC codes, Reed-Muller codes, Reed-Solomon codes, and BCH codes, etc. 
%Many of these codes were proved to be optimal in terms of coding rate or redundancy in some cases. 
Many of these codes were proved to be optimal in terms of redundancy in some settings. 
However, less is known about deletion and insertion errors, which are commonly referred to as \textit{synchronization} errors. %Nevertheless, there is an interesting fact about deletion and insertion errors is that if a code corrects deletion errors, it is also capable of correcting an equal number of combination of deletions and insertions. 
Nevertheless, the information-theorist Vladimir Levenshtein (1935-2017) proved an interesting fact about deletion and insertion errors already in the 1960's: if a code corrects deletion errors, it can also correct an equal number of combination of deletions and insertions (However, an efficient encoding/decoding algorithm for correcting deletions does not necessarily imply an efficient algorithm for correcting deletion and insertion errors). 
Moreover, a substitution error can be regarded as a deletion error followed by an insertion. Therefore, it is reasonable to focus on deletion errors, as correcting deletion errors implies correcting a combination of the three types of errors.  

%There are two scenarios considered for correcting deletion errors. One is the \textit{probabilistic} deletion channel, where a fraction of deletion errors occur randomly and the goal is to find the asymptotically optimal information rate (i.e., the amount of added redundancy relative to the data size), known as the channel capacity, such that on average, the channel input can be recovered, i.e., the deletion errors can be corrected. 
There are two scenarios for correcting deletion errors: 
\begin{enumerate}
    \item  The \textit{probabilistic} scenario, where a fraction of deletions occur randomly. The goal is to find the optimal information rate (i.e., the data size relative to the code length), known as the \textit{channel capacity}, such that the information could be recovered with high probability.
    \item %The \textit{adversarial} deletion channel, where the number of deletion errors is bounded and the goal is to correct any possible deletion errors satisfying the bound on the number. 
    The \textit{adversarial} scenario, where at most a certain number of deletions are caused by an adversary that wishes the reading process to fail. The goal is to find the minimal amount of added redundancy that guarantees successful reading in all cases.
    %to succeed in all cases. This is a preferable mode in cases where deletion correction is crucial in all cases, and the expected number of deletions is rather small.
\end{enumerate}

%In DNA storage applications, the number of deletions is normally quite small, and hence, the adversarial scenario is commonly preferred. Hence, it is more tempted to consider correcting adversarial deletion errors. 
DNA storage devices are commonly considered as an adversarial scenario, since the number of deletions is usually quite small, and the respective amount of redundancy can be optimized directly. Therefore, in this section we focus on \emph{adversarial deletion correcting codes}. Moreover, for simplicity we consider bits rather than nucleotides, however, similar statements can be made for nucleotides as well.
%As the coding rate approaches~$1$ for a small number of deletion errors and large code length, coding redundancy instead of coding rate is of more interest. In this section, we focus on adversarial deletion correcting codes. Moreover, we consider binary cases where codewords are sequences of bits. The binary codes can be extended to non-binary cases.
 %the coding rate approaches~$1$ for a small number of deletion errors and large code length, coding redundancy instead of coding rate is of more interest. In this section, we focus on adversarial deletion correcting codes. Moreover, we consider binary cases where codewords are sequences of bits. The binary codes can be extended to non-binary cases.

In contrast to multiple results about correcting substitution errors, there are not many efficient and well structured codes correcting deletion errors, and even the deletion channel capacity is still unknown except for cases where the deletion probability is small. 
One of the reasons for deletions and insertions being more difficult is that channels with substitution channels are \textit{memoryless}, i.e., different output bits are independent given the input, while deletion/insertion channels are not; one deletion affects all subsequent bits by shifting them one position to the left. Moreover, there is a symmetry in substitution errors that is not present in deletion or insertion errors. 

To see this symmetry, we introduce the notion of an \textit{error ball}, which is common in the analysis of error correcting codes. An error ball is the set of all possible sequences one can get after at most some number of errors occur in a given input sequence. If the type of error is substitution or deletion, we call the corresponding error ball a substitution ball or a deletion ball, respectively. 

For example, the substitution ball and the deletion ball of the input sequence $1001$ with at most~$1$ error are given by $\{1001,0001,1101,1011,1000\}$ and $\{1001,001,101,100\}$, respectively. The symmetry in substitution errors reflects the fact that the size of the substitution ball is independent of the input sequence; any other sequence of length~$4$ will have a substitution ball with~$5$ sequences. Moreover, the erroneous sequence is uniformly distributed over the substitution ball if the substitution indices are uniformly and randomly selected. % when enumerating all possible substitution error patterns. 
These two properties do not hold for deletion balls, since two deletion balls (for different input sequences) can have different sizes. In addition, the probabilities of getting different erroneous sequences in the deletion ball are different when the deletion indices are uniformly and randomly selected. The following example illustrates this crucial difference in symmetry.
\begin{example}
Consider the sequences $0000$ and $1010$. The substitution balls of $0000$ and $1010$, with at most a single substitution, are $\{0000,1000,0100,0010,0001\}$ and $\{1010,0010,1110,1000,1011\}$, respectively. Each set has $5$ elements and each element appears once under all possible error patterns. The deletion balls of $0000$ and $1011$ with at most a single deletion are given by $\{0000,000\}$ and $\{1011,011,111,101\}$, respectively. The numbers of elements in the two sets are different. In addition, after a single deletion in $1011$, the erroneous sequence becomes $011$ or $111$ if the first bit or the second bit is deleted, respectively. The erroneous sequence becomes $101$ if either the third or the fourth bit is deleted. Hence, it is more probable to obtain $101$ than $011$ or $111$ after a single deletion uniformly occurs in $1011$.
\end{example}
\subsection{How to efficiently correct deletions?}
One of the natural approaches to correct deletion errors is to use a \textit{repetition code}, which was presented in the introduction for substitution errors but also works well for deletion errors. This is undesirable due to high redundancy. Another potential approach is to borrow results from substitution correcting codes. The difficulty with this approach is that even two binary sequences with large \textit{Hamming distance} (i.e., have multiple different bits), which are resilient to substitution errors, can be ambiguous even under a single deletion. As an example, one can consider two sequences $1010101$ and $0101010$ that have Hamming distance~$7$ (i.e., all bits are different), and therefore one can identify the correct sequence between them under any~$3$ substitutions. However, these two sequences become indistinguishable in the case of even a single deletion, since deleting the first bit in one produces the same string~$010101$ as deleting the last bit in the other.
%given a channel output $010\ldots101$, which can be obtained from both sequences after a single deletion. 

How to correct a single deletion with low redundancy? One classic construction is the Varshamov-Tenengolt (VT) codes, defined by %\rcomment{Why not choose~$a=0$ and save a parameter? I changed it here and in the subsequent example.}
\begin{align}\label{eq:vtcode}
    \mathcal{C}_0=\{(c_1,\ldots,c_n):\textstyle\sum^n_{i=1}ic_i\equiv 0\bmod n+1\},
\end{align}
which corrects a single deletion. In words, VT codes show that one can protect any sequence from a single deletion by summing up the indices of all the $1$ entries in the sequence and taking a modulo of $n+1$. 
The following example demonstrates how this modulo summation scheme works.
\begin{example}
    Suppose the erroneous sequence $00110$ is obtained from some unknown sequence $c_1c_2c_3c_4c_5c_6$ after a single deletion. %In addition, we know that the modulo summation $\sum^n_{i=1}ic_i \bmod n+1$ is $4$. 
    To see what $c_1c_2c_3c_4c_5c_6$ is, enumerate all possible sequences of length~$6$ that might become~$00110$ after a single deletion, given by $000110$, $001010$, $001100$, $100110$, $010110$, $001110$, and $001101$. The respective modulo summations, i.e., the expressions~$\sum_{i=1}^6 ic_i\bmod(n+1)$, are given by $2$, $1$, $0$, $3$, $4$, and $5$, respectively, all of which are different. Therefore, the only sequence with modulo summation~$0$ is $001100$, and hence it must be the correct answer.
\end{example}
%The following is an example of VT code of length $6$
% \begin{example}
% Let $n=6$ and $a=0$ in \eqref{eq:vtcode}. Then, we have that
% \begin{align*}
%     \mathcal{C}_a=\{&000000,100001,010010,001100,001011,\\
%     &110011,101101,011110\}
% \end{align*}
% \end{example}
%It can also be verified that VT code is also a single erasure correcting code, i.e., it has Hamming distance at least $2$. In general, a code correcting a number of $t$ deletions is capable of correcting $t$ erasures, since erasure errors can be considered as deletion errors by knowing the indices where deletions occur. However, as we have shown in the previous example, the reverse does not hold.
%The modulo summation in the VT code is similar to the Reed-Solomon code correcting a single erasure (An erasure can be regarded as a deletion where the index of the deleted bit is known). 
The VT code has redundancy at most $\log_2 (n+1)$ bits, which is asymptotically optimal for a single deletion.  

Can we generalize the VT code to correct more than a single deletion with asymptotically optimal redundancy? This is a key question toward a good understanding of how to correct deletion errors, and turns out to be highly nontrivial. A counting argument by Vladimir Levenshtein showed that the optimal redundancy of codes of length $n$ correcting $t$ deletions lies between $t\log_2 n +o(\log_2 n)$ and $2t\log_2 n +o(\log_2 n)$ for constant $t$, where the notation $o(\log_2 n)$ means that~$\frac{o(\log_2 n)}{\log_2 n}$ approaches $0$ as $n$ goes to infinity. For larger $t$, the optimal redundancy is linear in $t\log_2(\frac{n}{t})$. These arguments only prove that a code exists, without finding it explicitly. 
%Note that the length $n$ code correcting $t$ deletions with such redundancy is only existential and constructed by exhaustive search, which is intractable. 
Finding the respective codes explicitly with comparable redundancy has been puzzling for decades even for the case $t=2$.

%%% I removed the reference to RS codes, I'm not sure it adds much. -Netanel. %%%
% Inspired by the fact that the modulo summation in VT codes has a similar structure to the RS code correcting a single erasure, (An erasure can be regarded as a deletion where the index of the deleted bit is known) it is natural to think if the multiple-erasure correcting RS code, which takes the summation of higher order weights, work for correcting multiple deletion errors. More specifically, given the modulo sums $\sum^n_{i=1}i^pc_i$ of powers of indices of order $p=0$ up to some positive integer, is it possible to recover $(c_1,\ldots,c_n)$ from multiple deletions? This is not clear even for two deletions. Counterexamples were found showing that knowing the sums $\sum^n_{i=1}i^pc_i$ of order $p=0$ up to $p=4$ does not guarantee successful correction. 
Inspired by the modulo summation in VT codes, researchers wondered if \textit{higher-power} summation might be useful for correcting multiple deletions with optimal redundancy. Specifically, given the modulo sums $\sum^n_{i=1}i^pc_i$ for~$p$ from~$0$ up to some positive integer, is it possible to recover $(c_1,\ldots,c_n)$ from multiple deletions? This is a challenging question even for two deletions, and unfortunately, counterexamples were found showing that knowing the sums $\sum^n_{i=1}i^pc_i$ for~$p$'s from~$0$ up to $p=4$ does not guarantee successful correction.

To generalize the idea of using weighted modulo sum for correcting multiple deletions, one can use weights that are exponential in indices in the weighted modulo sum. However, due to the exponential weights, such generalization requires redundancy that is linear in the code length to correct even two deletions, in contrast to the optimal redundancy which is a logarithm in the code length for correcting a constant number of deletions. %Though algebraic approaches such as weighted modulo sum parity checks achieved tremendous success in correcting substitution errors, it is illusive whether they work for deletion errors.

Another brilliant idea for correcting multiple deletions is to use a concatenated code, that has a two-level structure of an inner code and an outer code. %The idea of concatenated code was introduced by Forney and then used by Schulman and Zuckerman to propose the first asymptotically good codes correcting multiple adversarial deletions. The codes correct a (small) fraction of $n$ (the code length)  deletions with redundancy linear in $n$.  
%In their deletion correcting codes, 
Specifically, the codewords are separated into blocks. The inner code protects each block from deletions and is constructed by using exhaustive search (i.e., finding the best code by traversing all codes using a computer). %(Recall that the existence of a deletion correcting code with small redundancy can be constructed by brute force search.) 
The outer code treats each block as a symbol and uses a substitution correcting code to correct blocks in case the inner code fails in some blocks. 
Note that the brute force search to construct the inner code is tractable when the block size is small (specifically, a logarithm of the code length). The concatenated code approach reduces the problem of correcting deletions in a long sequence to that of correcting deletions in short sequences, by using the well-constructed substitution correcting codes. 
%Pursuing the direction of using concatenated codes, Guruswami and Wang improved the result by Schulman and Zuckerman for correcting a small fraction of deletions. Guruswami and Li proposed codes correcting $\frac{1}{3}$ fraction of deletions with redundancy linear in the code length. 
Using concatenated codes, it is possible to correct a number of deletions which is linear in the code length, with redundancy that is also linear in the code length. This is asymptotically optimal based on the bounds that were mentioned above. 

What about correcting a small number, say a constant, of deletions, a regime that is of interest in DNA storage since the number of deletions is small and the code length is moderately sized? As discussed previously, the optimal redundancy should be asymptotically between $t\log_2 n$ and $2t\log_2 n$ where $t$ is the number of deletions.
In the following, we discuss codes with close to optimal redundancy, using a fundamentally different idea from the concatenated code construction. Specifically, we discuss a generalization of the VT codes using an algebraic approach.  Recall that the higher order weighted sum $\sum^n_{i=1}i^pc_i$, a natural generalization of the VT codes, is not guaranteed to provide a code correcting even two deletions. However, a similar higher order weighted sum is capable of correcting deletions for constrained sequences. 

We illustrate the idea for the case of two deletions. In this case, an interesting observation is that if the codewords are sequences with at least a single $0$ between any two $1$'s, then any codeword $(c_1,\ldots,c_n)$ can be protected from two deletions by providing the sums $\sum^n_{i=1}(\sum^i_{j=1}j^p)c_i\bmod 2n^{p+1}$ for $p=0,1,2$. %The following example illustrates this observation.
\begin{example}
    Consider the sequence $101001$ of length~$6$, where any two $1$'s are separated by at least one $0$. Its weighted modulo sums are %given by a vector 
    \begin{align*}
    \textstyle\sum^6_{i=1}(\sum^i_{j=1}j^0)c_i\bmod 2\cdot6^{0+1}&=3\\
    \textstyle\sum^6_{i=1}(\sum^i_{j=1}j^1)c_i\bmod 2\cdot6^{1+1}&=28\\
    \textstyle\sum^6_{i=1}(\sum^i_{j=1}j^2)c_i\bmod 2\cdot6^{2+1}&=106.
    %=(3,28,106).
    \end{align*}
    Suppose two deletions occur in $101001$, resulting in $1001$. We list all possible sequences that satisfy: (1) can result in $1001$ after two deletions; and (2) there is at least one $0$ bit between any two $1$ bits. These sequences are given by $001001,010001,010010,100001,100010,100100,10100,$ $1,010101,$ and~$100101$. It can be verified that only $101001$ has the weighted modulo sums $3,28,106$ as above.
\end{example}

But how do we guarantee at least one~$0$ between any two~$1$'s? To this end, let us define an \textit{indicator vector} $\mathbbm{1}_{10}(c_1,\ldots,c_n)$ of length $n$ for a sequence $(c_1,\ldots,c_n)$ as follows. The~$i$'th bit of~$\mathbbm{1}_{10}(c_1,\ldots,c_n)$ is
\begin{align*}
    \mathbbm{1}_{10}(c_1,\ldots,c_n)_i=\begin{cases}
			1 & \mbox{if }c_i=1,\mbox{ and }c_{i+1}=0\\
			0 &\mbox{otherwise}.
		\end{cases},
\end{align*}
for $i\in\{1,\ldots,n\}$, where it is assumed that $c_{n+1}=1$. 
Note that by definition, for any binary sequence $(c_1,\ldots,c_n)$, there is at least one $0$ between any two $1$'s in $\mathbbm{1}_{10}(c_1,\ldots,c_n)$: it is impossible to have to consecutive~$1$'s, since it would imply both that~$c_i=1$ and~$c_{i+1}=0$, and that~$c_{i+1}=1$ and~$c_{i+2}=0$, and clearly~$c_{i+1}$ cannot simultaneously be a~$0$ and a~$1$.

Therefore, for any sequence $(c_1,\ldots,c_n)$, the indicator vector $\mathbbm{1}_{10}(c_1,\ldots,c_n)$ can be protected from two deletions given the weighted modulo sums $\sum^n_{i=1}(\sum^i_{j=1}j^p)\mathbbm{1}_{10}(c_1,\ldots,c_n)_i\bmod 2n^{p+1}$ for $p=0,1,2$. Then, it suffices to protect the modulo sums $\sum^n_{i=1}(\sum^i_{j=1}j^p)\mathbbm{1}_{10}(c_1,\ldots,c_n)_i\bmod 2n^{p+1}$, $p=0,1,2$ by using a very short repetition code, in order to recover the vector $\mathbbm{1}_{10}(c_1,\ldots,c_n)$. After recovering the vector $\mathbbm{1}_{10}(c_1,\ldots,c_n)$, one can protect a similarly defined indicator vector $\mathbbm{1}_{01}(c_1,\ldots,c_n)$. Finally, 
the sequence $(c_1,\ldots,c_n)$ can be recovered from indicator vectors $\mathbbm{1}_{10}(c_1,\ldots,c_n)$ and $\mathbbm{1}_{01}(c_1,\ldots,c_n)$. %\textcolor{red}{[The last sentence is unclear, can we rephrase?]}%Note that the two deletion correcting codes generalize the VT codes from an algebraic approach.

This approach extends to any constant number $t$ of deletions by generalizing the observation: For any $t$, if the codewords are sequences that have at least $t-1$ $0$'s between any two $1$'s, then the codewords can be protected from $t$ deletions by using the weighted modulo sum $\sum^n_{i=1}(\sum^i_{j=1}j^p)c_i\bmod 3tn^{p+1}$ for $p=0,1,\ldots,6t$. Similarly, one can define indicator vectors such that the indicator vector of a sequence has at least $t-1$ $0$'s between any two $1$'s. The resulting redundancy is then $4t\log_2 n +o(\log_2 n)$, which is asymptotically at most $4$ times the optimal. %By combining the constructions of Sima, Gabrys, and Bruck and the VT codes, Song, Polyanskii, Cai, and He improved the redundancy to $(4t-1)\log_2 n +o(\log_2 n)$, which is currently the best known redundancy result for correcting $t$ deletions.

Despite the progress on codes correcting $t$ deletions, several problems remain open. How to construct minimal-redundancy deletion codes which can also be decoded efficiently? 
%that are both redundancy and computationally efficient? 
How to approach or improve the existential redundancy bound? How to efficiently correct a combination of deletions and substitutions? %For the third question, Song, Polyanskii, Cai, and He proposed a code with redundancy $4t\log_2 n + 3s\log_2 n +o(\log_2 n)$ for correcting a combination of $t$ deletions and $s$ substitutions, based on the BCH codes and codes of Sima, Gabrys, and Bruck. 
The third problem is crucial in DNA storage applications as errors are normally a combination of deletions, insertions, and substitutions. Though the problem 
was investigated and codes combining the deletion codes above and the substitution codes were proposed, 
are there more redundancy efficient methods?

\begin{figure}[t]
\centering
\includegraphics[width=1\linewidth]{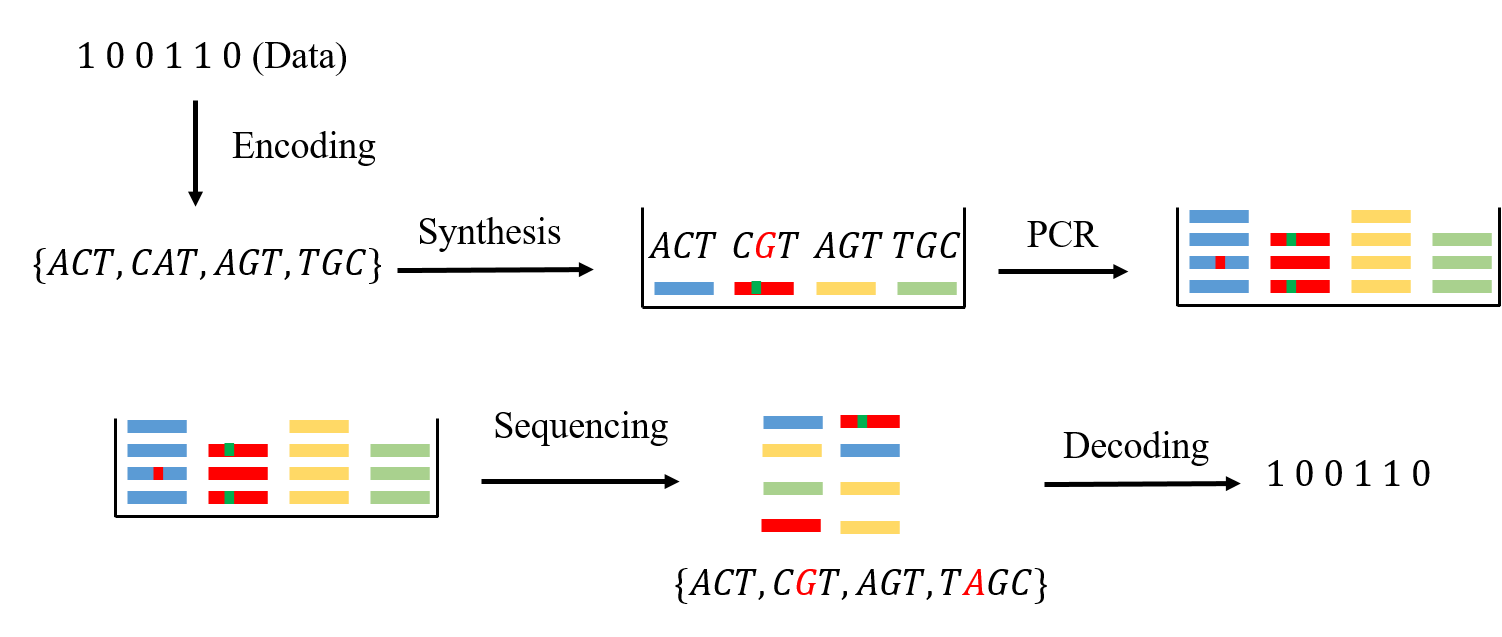}
\caption{An illustration of the processes in a typical in-vitro DNA storage system. The data $(1,0,0,1,1,0)$ is encoded into a set of short sequences $\{ACT,CAT,AGT,TGC\}$ of nucleotides. Errors occur during the synthesis of the short sequences, turning the nucleotide $A$ in  $CAT$ to $G$. The synthesized sequences, including the erroneous sequence "$CGT$", are amplified in the PCR process, generating multiple noisy copies of the synthesized sequences. During the sequencing process, these noisy copies are read and clustered such that each cluster consists of noisy copies of a synthesized sequence. Then, an estimate $\{ACT,CGT,AGT,TAGC\}$ of the synthesized sequences is obtained from the clusters of noisy copies, where $TAGC$ is an erroneous estimate of $TGC$. Finally, the estimated set of synthesized sequences is decoded into data.}
\label{fig:examplepartition}
\end{figure}

\section{Sliced Channel}
\label{section:sliced}

One feature that fundamentally distinguishes DNA storage from traditional storage is that in
traditional systems, the codeword is a single long sequence, whereas in DNA storage, the codeword is a set of unordered short sequences. 

To see this, we briefly describe the workflow of DNA storage systems, shown in Fig \ref{fig:examplepartition}. As mentioned earlier, in DNA storage information is encoded into sequences of four letters, $\sA, \sC, \sG, \sT$, which are synthesized into the respective DNA molecules.
%This is the synthesis process that corresponds to the writing of the data. 
The synthesized DNA molecules are then placed in a solution inside a vial. In the reading phase, the sequences are amplified by a process called Polymerase Chain Reaction (PCR), which generates many more copies of the nucleotide sequences in the vial. The copies are then sampled and read through a sequencing process, producing many potentially erroneous copies of the sequences that were originally synthesized. By using clustering and reconstruction algorithms, the copies generated from the same sequence are clustered, and the corresponding sequence is reconstructed. Finally, the reconstructed sequences are decoded to retrieve the data. Due to technological constraints in the above processes, only short DNA molecules ($\approx 100$ nucleotides) can be synthesized and sequenced, meaning that information can only be encoded into a collection of short sequences. Moreover, the DNA molecules stored in the same vial are unordered; they all float in the same solution without any knowledge regarding which comes prior.
%pool do not have an identifier so that a DNA molecule stores no information beyond its nucleotide sequence. Therefore, the collection of nucleotide sequences are unordered. 

Note that errors occur in the collection of unordered short nucleotide sequences. This gives rise to the question of how to correct errors when the codeword is ``sliced'' into multiple unordered pieces. This brings new aspects to classic error correction setups, where the information is encoded into a codeword that is a single sequence, and retrieved from a noisy copy of that sequence. In the context of coding for DNA storage, the codeword is sliced into multiple unordered pieces, normally of equal lengths, which are presented both noisy and unordered to the decoder. 
%and the noisy copy of the codeword becomes multiple unordered noisy copies of different codeword pieces. 

In the sliced codeword setting, we may either think of a codeword as a single sequence that it then sliced into multiple unordered pieces, or a priori consider the codeword as the set of those pieces; both approaches are equivalent.
%it makes no difference whether the codeword is a single sequence or a set of short sequences that are sliced pieces of a single sequence. Hence
In this paper, we choose the latter, i.e., we assume that the codeword is a set of $M$ short sequences, each of length $L$.  In existing DNA storage systems, $L$ is of the order of magnitude $100$ and $M$ is of the order of magnitude $10^4$ to $10^9$, based on the size of the data.

In DNA storage, the types of errors include: (1) deletions, insertions, and substitutions, that occur in either of the sequences, and were discussed in Section~\ref{sec:delcodes}. (2) sequence loss due to the fact that some DNA molecules may not be sampled during the sequencing process. As a result, the sequence contained in the DNA molecule is missing. 
The following is an example of coding over a set of short binary sequences.
\begin{example}
    Assume that the given data is encoded to~$M=4$ sequences of length~$L=6$, say $\{110001,100100,101010,111111\}$, which are placed together in a vial.
    %Let $M=4$, $L=6$, and a codeword be a set of $M$ sequences $\{110001,100100,101010,111111\}$ of length $L$. 
    Then, noisy copies of the codeword can be $\{100000,1100001,10101\}$, after deletion, insertion, substitution errors, and sequence loss that occur in any sequence in the set. Note that the identity of a noisy copy is not known. For example, the noisy copies $100000$, $1100001$, $10101$ can be obtained from $110001$, $100100$, $101010$, respectively, or from $100100$, $110001$, and $111111$, respectively.
\end{example}
The setting of coding over a set of sequences can be considered as a generalization of the classic setting of coding over a single sequence, where the set contains only a single sequence and there is no sequence loss. %This general setting may also find applications 
Also, a similar ordering issue often arises in network packet transmission, where due to varied network delays and changes in routing, packets might arrive not in the same order they were sent.
%where some packets are lost and the order of packets arriving at the receiver may be different from the order of packets sent from the encoder, due to different routing paths.

To understand the problem of how to handle unordered sets of sequences, we focus on the basic setting where the codeword is a set of $M$ different binary sequences of length $L$, and focus on substitution errors; more complex settings follow similar ideas. %Error correcting codes for more general settings can be constructed based on the approach for this basic setting. 
For example, one can transform a code that uses~$\{0,1\}$ to one which uses $\sA, \sC, \sG, \sT$ using the mapping mentioned in the introduction.
%binary code to a $4$-ary code, suited for nucleotide sequences with four bases, by simply merging two bits into a single $4$-ary symbol. 
Further, to correct deletion and insertion errors, one can combine the codes for this basic setting and the deletion codes discussed in Section \ref{sec:delcodes}. To combat sequence loss, it is possible to add an ``outer'' code, such as the Reed-Solomon code. % on top without sequence loss, such as RS code, to correct the lost sequences.   

Intuitively, correcting errors in the sliced codeword setting (over a set of sequences) is more difficult than correcting errors in a single sequence, since in the former setting, the information about the index of each sliced piece is lost. 
One natural way to correct errors in the set of sequences is to use error correction codes to protect each sequence independently. This is efficient when each sequence roughly has the same amount of errors. In the case when some sequences have no errors, or some sequences have much more errors than average, the method may be inefficient since one has to protect every sequence from the largest number of errors possible.

To deal with the loss of the index information of each unordered sequence, another natural (and possibly the most common) approach is to use extra redundancy to index each sequence. That is, in each sequence, dedicate the first $\log_2 M$ (out of $L$) bits to record the index among the total $M$ sequences. This gives an order to the sequences based on their indices, and reduces the problem of coding over an unordered set of sequences to that of coding over a single sequence. 

It can be shown that the simple index-based scheme asymptotically approaches the best information rate, i.e., the channel capacity, 
for coding over an unordered set of sequences. More specifically, index-based schemes achieve the asymptotically optimal information rate in probabilistic settings where a fraction of sequence loss and substitution errors randomly occur.  %Suppose there is no error, then the information rate of the indexing scheme, i.e., the number of bits which store information divided by the total number of bits in a codeword, is given by $\frac{(L-\log_2 M)M}{ML}=1-\frac{\log_2 M}{L}$. Note that intrinsically, the information rate due to the lost order is at most $\frac{\log_2 \binom{2^L}{M}}{ML}$, since the total number of codewords is $\binom{2^L}{M}$. Therefore, the information rate loss due to indexing is at most $\frac{\log_2 \binom{2^L}{M}-(L-\log_2 M)M}{ML}$, which equals $\frac{o(L)}{L}$ whenever $M\le 2^{\beta L}$ for some constant $\beta<1$. This rate loss is negligible as $L$ becomes large, meaning that the indexing scheme is asymptotically rate optimal when there are no errors. With this observation, it can be shown that indexing schemes are asymptotically rate optimal when only sequence loss is possible. The asymptotic optimality of indexing schemes extends to channels with both sequence loss and substitution errors. 
Partly for this reason, indexing schemes are used in most of the recent DNA storage experiments, where extra bases are dedicated to index each sequence, and Reed-Solomon codes are used to correct errors in the bases. In addition, many code constructions were proposed based on the indexing schemes.

One of the problems which need to be addressed for indexing schemes is to protect against errors in the index. 
One way is to encode the indices such that they are far from each other in Hamming distance (i.e., have many distinct bits). In this way, the indices are more robust to substitution errors. However, it requires more redundancy in the indices. 
To resolve this issue, another approach is to use data to protect errors in indices. When the Hamming distance between two indices is small, meaning that they are ambiguous under substitution errors, it is required the data in the corresponding two sequences have a large enough Hamming distance. With this constraint, the decoder can distinguish two sequences based on their data, if it fails to decode their indices. %Ignoring the redundancy caused by indexing, it is possible to achieve asymptotically order-wise optimal redundancy $t\log_2 M$, where $s$ is the number of erroneous sequences and $\epsilon$ is the number of errors in each erroneous sequence.

%Besides the $\log_2 M$-bits index in each sequence, an anchor of length close to $\log_2 M$ is added to each sequence, such that if the indices of two sequences have small Hamming distance and indistinguishable, their anchors have enough Hamming distance to distinguish them. The construction of such anchors was proposed by Shinkar, Yaakobi, Lenz, and Wachter-Zeh, who showed that it costs at most a single bit redundancy to protect the indices by using anchors. Note that the anchors also encodes information. Ignoring the redundancy caused by indexing, the anchor-based scheme was shown to achieve asymptotically order-wise optimal redundancy $2s\epsilon\log_2 ML+2s\log_2 M$, where $s$ is the number of erroneous sequences and $\epsilon$ is the number of errors in each erroneous sequence. Another code construction protecting indices from errors was proposed by Song, Cai, and Immink, where indices were replaced by sequences with lower bounded mutual Hamming distance.  

While index-based schemes achieve asymptotically the optimal information rate in probabilistic settings, how do they perform in deterministic settings? In deterministic settings, the number of errors is bounded and zero-error decoding is required. 
When the number of errors is not large, it is reasonable to look at the redundancy, rather than the information rate of a scheme, which approaches one with a small number of errors. 
Note that different from information rate, which measures the ratio between the amount of information and the number of symbols used to store the information, redundancy measures the difference between the two. 
%Therefore, asymptotically redundancy optimal schemes are asymptotically information rate optimal but not vice versa. 
But how should we define ``redundancy'' in the unordered sequence setting?
%is it optimal in redundancy, which is analogous to the second order term in the information rate? 
To study this question, we define the redundancy of a code as $\log_2 \binom{2^L}{M}-
\log_2 |\mathcal{C}|$, where $|\mathcal{C}|$ is the size of the code (i.e., number of codewords) and $\log_2 |\mathcal{C}|$ is the number of information bits the code can represent. This definition measures how many extra bits are needed for error correction, and would be zero in the case of no errors. 
%It was (to the best of our knowledge) introduced in the work of Lenz, Siegel, Wachter-Zeh, and Yaakobi. 
Under this definition, the extra redundancy needed for indexing in an index-based scheme is at least $\log_2 \binom{2^L}{M}-
M(L-\log_2 M)$ (which is linear in $M$) even when there are no errors. 

%How does the linear in $M$ redundancy compare to the optimal one? 
Can one use less redundancy than that? Using counting arguments, %a Gilbert-Varshamov type bound can be obtained, which suggests that 
one can show that the optimal redundancy for correcting a total number of $t$ substitution errors across all $M$ unordered sequences of length $L$ is at most $2t\log_2 (ML)+o(\log_2 (ML))$ for small $t$ (e.g., a constant), and at most linear in $t\log_2 (\frac{ML}{t})$
for large $t$ (e.g., a fraction of $ML$). In addition, the optimal redundancy is at least $t\log_2 (ML)+o(\log_2 (ML))$ for small $t$, and at least linear in $t\log_2 (\frac{ML}{t})$ for large $t$. The upper and lower bounds are order-wise the same. 
Note that in the classic setting of correcting $t$ substitution errors over a single sequence of length $ML$ (which is equivalent to $M$ ordered sequences of length $L$), similar counting arguments 
%Gilbert-Varshamov upper bound on the redundancy 
also yield $2t\log_2 (ML)+o(\log_2 (ML))$ for small $t$, and linear in $t\log_2 (\frac{ML}{t})$ for large $t$. In addition, the lower bound on redundancy is order-wise the same as the upper bound. This result has a surprising implication: 
%Though it seems more difficult to correct errors over a set of sequences than over a single sequence due to the loss of index information, the extra redundancy needed for error correction is almost the same for both cases. 
\textbf{it costs almost the same amount of redundancy to correct errors over a set of unordered and ordered sequences!} This is highly surprising, since the unordered case is intuitively more complex.

%We now compare the Gilbert-Varshamov type redundancy ($2t\log_2 (ML)+o(\log_2 (ML))$) to that (linear in $M$) of the index-based schemes. 
We now compare the redundancy bound $2t\log_2 (ML)+o(\log_2 (ML))$ to that of the index-based schemes presented earlier, which is linear in $M$. Since $M$ is much larger than $L$, the redundancy in the index-based schemes is much larger than bound $2t\log_2 (ML)+o(\log_2 (ML))$ whenever $t=o(\frac{M}{\log_2(ML)})$. %\rcomment{While the parameter regime $t=o(\frac{M}{\log_2(ML)})$ may not fit DNA storage systems under current technologies, in which~$t$ is usually larger than that. However, 
%as read/write errors occur more often in current technologies. But 
%the ideas discussed in this section may be applied in other scenarios [I'm not sure why we need this sentence, it's a little demotivating. Instead I suggest:]} 
In what follows we describe ideas that close this gap, i.e., provide codes correcting substitution errors with almost optimal redundancy, and thus are better than index-based schemes for this many substitution errors.

%In the following, we discuss codes correcting substitution errors with order-wise the same as the optimal redundancy, and thus are better than index-based schemes for small number of errors. 
\begin{figure}[t]
\centering
\includegraphics[width=1\linewidth]{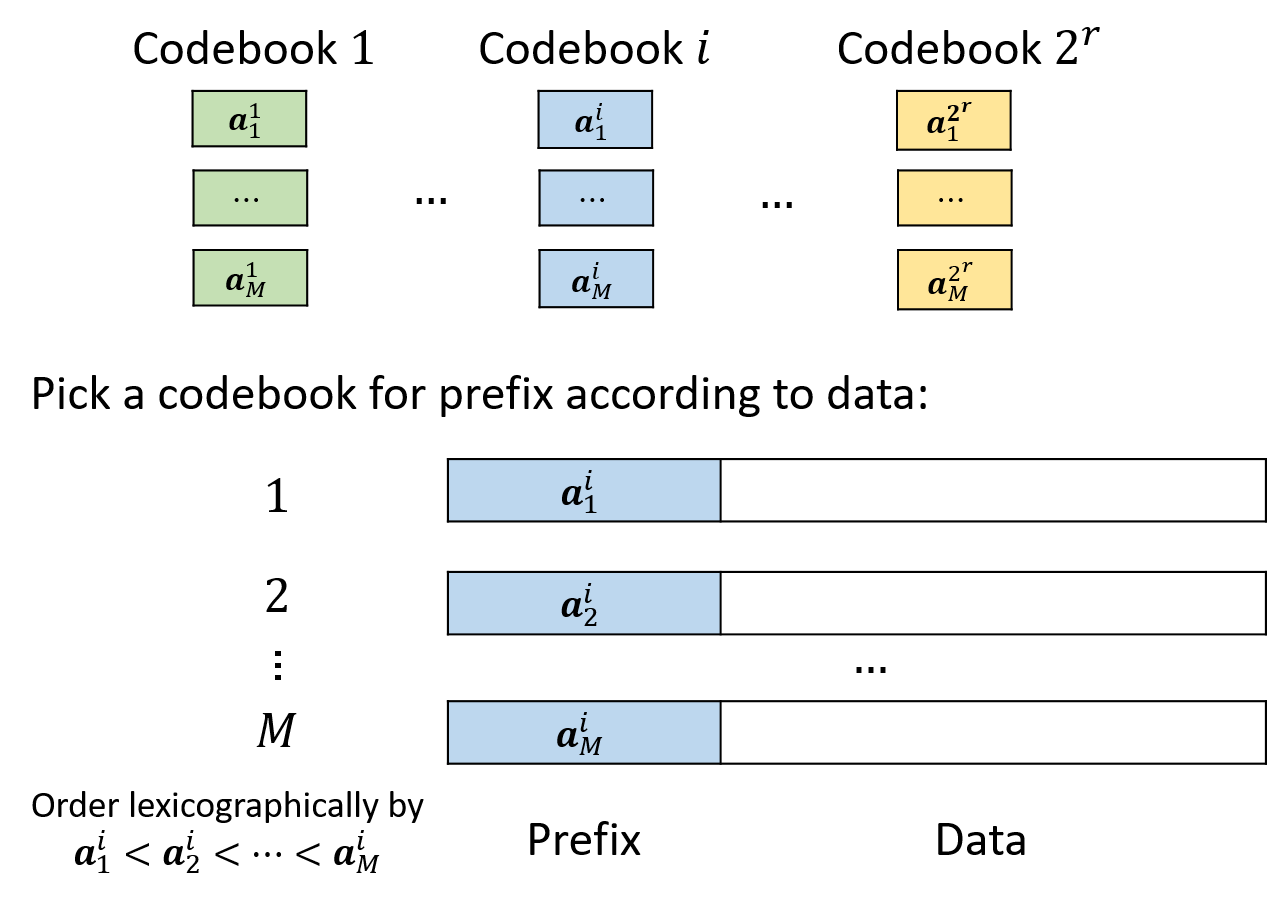}
\caption{An illustration of codes that use data for indexing.}
\label{fig:slicedcoding}
\end{figure}
The idea is to use the data itself for the purpose of indexing, or alternatively, encoding data inside the index. %The statistical correlation between data and the identity of the noisy copies was also noticed by Shomorony and Heckel in the proof that indexing schemes achieve channel capacity. Yet, it was not used for coding since indexing schemes already achieve the channel capacity. 
Specifically, we use the lexicographic order of the data for indexing, that is, the prefix in each sequence is used for indexing, while also containing data. %The following is an example showing how sequences are indexed.
 \begin{example}
     Let the codeword be $\{1001101,0101100,1010001,0001001\}$, where the first three bits are the prefix in each sequence used for indexing. Then one can order the set of sequences in the codeword by $0001001,0101100,1001101,1010001$ in ascending lexicographic order of the prefixes.
 \end{example}
Using prefixes for indexing is similar to index-based schemes. Yet, unlike index-based schemes, the prefixes also encode information. In order to make the indexing of the prefixes robust from substitution errors, the collection of prefixes in all sequences constitutes a code with %a lower bounded 
large minimum Hamming distance. This is similar to protecting the indices from errors in the index-based schemes. The difference is that the indexing prefixes here encode information. Information is encoded into prefixes through different choices of the codes, as shown in Fig. \ref{fig:slicedcoding}. The construction of codes for the prefixes can be done using a greedy algorithm, so that the prefixes in the code are generated bit-by-bit. %When there are no errors, this greedy algorithm reduces to the \textit{combinatorial numbering algorithm}, which maps an integer to a set of different elements \rcomment{[Even I don't remember what that means. :) Rephrase?]}. 

The scheme of using data to index avoids the index bits, and achieves redundancy that is linear in $t\log_2 (ML)$, i.e., almost optimal. One can also combine it with the deletion correcting codes discussed in Section~\ref{sec:delcodes}, and enable deletion and/or insertion correction as well. %In addition, the idea of using lexicographic order of data for indexing also applies to other settings. 

Finally, some questions about the unordered setting remain unanswered: 
%The questions remain for the problem of coding over set of multiple sequences include: 
What is the optimal redundancy for a \textit{large} number of errors? How to construct efficient codes that achieve order-wise optimal redundancy?
%Similar idea of using lexicographic order of data for indexing was also considered in the codes proposed by Wei and Schwartz, for a more general setting where sequence loss and deletion, insertion, and substitution errors are considered. %The codes of Wei and Schwartz achieve redundancy logarithmic in $ML$, which improves previous results of Lenz, Siegel, Wachter, and Yaakobi, and is order-wise optimal. 

%incurs high redundancy since in current DNA storage systems, the length of each sequence, denoted by $L$, is of the order of magnitude $100$, while the number of sequences, denoted by $M$, is of the order of magnitude $10^4$ to $10^9$ based on the size of the data. Adding redundancy to each sequence results in $M$ times the redundancy needed to protect a single sequence.  

\section{Duplication}\label{section:duplication}

Unlike previous sections, this section is motivated by storing information in the DNA of living organisms. The process involves synthesizing DNA sequences which are then inserted into the DNA of living organisms. We can then sequence the DNA extracted from these organisms, or more likely, their descendants, to read the information. Thus, the DNA storage channel in this case corrupts data not only due to synthesis and sequencing errors, but also due to naturally occurring biological processes that mutate the DNA.

We now focus on the errors (mutations) introduced by biological processes. It is well known that when cells divide, the genetic material is replicated. However, the DNA replication process is not without noise, and the resulting copy may be corrupted by several error types. These include substitution, where a base is replaced by another (point mutation), as well as insertions and deletions of blocks. These types of errors have been studied to various extents by existing literature. Another type of error is duplication, whereby a copy of a substring of the DNA is inserted. Duplications accumulate over time, and it has been found that the majority of human DNA is duplicated. Since this error type is rarely found in electronic communication, it has not been studied in the coding theory community, and in what follows, we focus on it solely.

What kind of duplications are possible? Several biological mechanisms are known to create duplications in the process of replicating the DNA. We illustrate a few here. Perhaps the simplest one (though not necessarily the most common) is \emph{tandem duplication}. This mutation process takes a substring and inserts a duplicate of it immediately after its original location, for example,
\[ \mathsf{ACCTAGGA} \der \mathsf{AC\underline{CTA}\overline{CTA}GGA} \quad\text{(tandem),}\]
where the underlined part is the substring being duplicated, and the overlined part is the inserted duplication. In \emph{interspersed duplication}, the duplicated part is inserted anywhere in the sequence, for example
\[ \mathsf{ACCTAGGA} \der \mathsf{AC\underline{CTA}GG\overline{CTA}A} \quad\text{(interspersed).}\]
Another duplication process, called \emph{reverse-complement duplication} (r.c.~duplication), takes the substring to duplicate, and inserts a reversed and complemented duplicate of it immediately after its location, for example
\[ \mathsf{ACCTAGGA} \der \mathsf{AC\underline{CTA}\overline{TAG}GGA} \quad\text{(r.c.).}\]
Here we use the Watson-Crick base pairing, making $\mathsf{A}$ and $\mathsf{T}$ complements of each other, and similarly, $\mathsf{C}$ and $\mathsf{G}$. All of the processes mentioned above are the result of known biological mutation processes whose mechanisms we understand. For the sake of mathematical simplicity, and to better illustrate the intricacies of duplication processes, we introduce an artificial duplication process called \emph{end duplication} which inserts the duplicated part at the end of the sequence, for example,
\[ \mathsf{ACCTAGGA} \der \mathsf{AC\underline{CTA}GGA\overline{CTA}} \quad\text{(end).}\]
As a final note on duplication processes, we emphasize that following a duplication process, another may occur, perhaps of a different type, and perhaps of a different length. Thus, over time, duplications accumulate, like layers of an onion. A naive inspection of a DNA sequence may only reveal the outer layer, namely, the last duplications made, and only after removing those, older occurrences become visible.

We can now formalize the description of the duplication channel. We shall be working over some finite alphabet $\Sigma$ (which in the case of DNA molecules will be $\Sigma=\{\mathsf{A},\mathsf{C},\mathsf{G},\mathsf{T}\}$). We store a sequence $x\in\Sigma^*$ where $\Sigma^*$ is the set of all finite length sequences over $\Sigma$. The channel then applies any number of duplications, resulting in a string $y\in\Sigma^*$. We denote this process as $x\der^* y$. The set of all possible mutated outcomes, given that $x$ was stored, is called the \emph{descendant cone of $x$}, and is denoted by $D^*(x)$. Conversely, the set of all possible strings that may be mutated by the channel into $y$ is called the \emph{ancestor cone of $y$}, and is denoted by $A^*(y)$.

When faced with such a channel, our goal is to construct error-correcting codes that can undo the duplications and recover the original stored sequence. General coding-theoretic principles guide us to define an error-correcting code as a set of sequences $C\subseteq\Sigma^n$, whose descendant cones are disjoint, namely, for any $c,c'\in C$, $D^*(c)\cap D^*(c')=\emptyset$. Thus, any corrupted sequence belongs to a single descendant cone of a valid codeword, and the decoding process simply outputs that codeword in response.

Many questions arise: How do we find a good error-correcting code? What makes a good error-correcting code? What is the best possible? How do we encode, and how do we decode? How does the answer depend on the type and parameters of the duplication processes? In what follows we briefly outline partial answers to these questions, and along the way, uncover connections to other motivating problems. 

\subsection{Know thy enemy -- Understanding descendant cones}

The first property of interest, when studying descendant cones, is knowing their size. Since our ultimate goal is constructing error-correcting codes, which are equivalent to packing descendant cones without overlap, finding their size may help us bound the parameters of such codes. The number of strings in any descendant cone is obviously infinite, and thus, we do not measure their size but rather the rate at which they grow with each mutation step. This property is called the \emph{capacity}, and sometimes the \emph{combinatorial entropy}.

Formally speaking, to compute the capacity of the descendant cone of $x\in\Sigma^*$, the definition calls for counting the number of descendants of length $n$, i.e., $\abs{D^*(x)\cap\Sigma^n}$. Taking $\log_2$ of this number and dividing by $n$ gives us the exponential growth rate we are after. Thus,
\[ \ccap(x) = \limsup_{n\to\infty} \frac{1}{n}\log_2\abs*{D^*(x)\cap\Sigma^n}.\]
A large capacity indicates a fast growing descendant cone, and similarly, a small capacity indicates slow growth. Packing fast growing descendant cones may be more difficult, resulting in smaller, less efficient, error-correcting codes.

The capacity may obviously depend on the alphabet size, the starting sequence $x$, and the duplication rules. To illustrate the subtleties of the latter, fix an alphabet $\Sigma$ and a starting sequence $x$. First consider the end-duplication system, in which each mutation copies a fixed-length substring of length $k$ to the end. It has been shown that this has full capacity, i.e., $\ccap^{\mathrm{end}}_k(x)=\log_2\abs{\Sigma}$, 
which is the highest possible value the capacity may have, indicating the highest possible growth rate for a descendant cone. We now tweak a single parameter -- instead of end duplication, we consider tandem duplication, namely the duplicated sequence of fixed length $k$ is inserted immediately after its original position. With this minute change, the capacity vanishes completely, i.e., $\ccap^{\mathrm{tan}}_k(x)=0$, 
indicating sub-exponential growth of the descendant cone.

One might argue that the capacity is perhaps too harsh: it takes into account what is \emph{possible}, where instead it should take into account what is \emph{probable}. We can describe the mutations as a stochastic process. We start with the initial sequence $x\in\Sigma^*$, and then at each round, a randomly chosen duplication rule (duplicating substring of fixed length $k$) is applied to a randomly selected position. We can set, to our liking, the distributions from which the duplication rule and the location are chosen. We denote the resulting sequence after $n$ mutations by $S_n(x)$, and observe that it is a random variable. We can then define the \emph{entropy} of $S_n(x)$ as
\begin{multline*}
H(S_n(x))= \\
-\sum_{w\in\Sigma^*}\Pr(S_n(x)=w)\log_2\Pr(S_n(x)=w).
\end{multline*}
With this, the \emph{entropy rate} of the entire system
\[ h(S(x))= \limsup_{n\to\infty}\frac{1}{n}H(S_n(x)).\]
Loosely speaking, $h(S(x))$ measures the amount of information generated by an application of a random duplication rule. Using standard information-theoretic arguments one can show that the capacity bounds the entropy rate from above, namely,
\[ h(S(x)) \leq \ccap(x).\]

\begin{figure}
    \centering
    \begin{overpic}[scale=0.3] % Change 0.6 to 0.3 for two columns
    {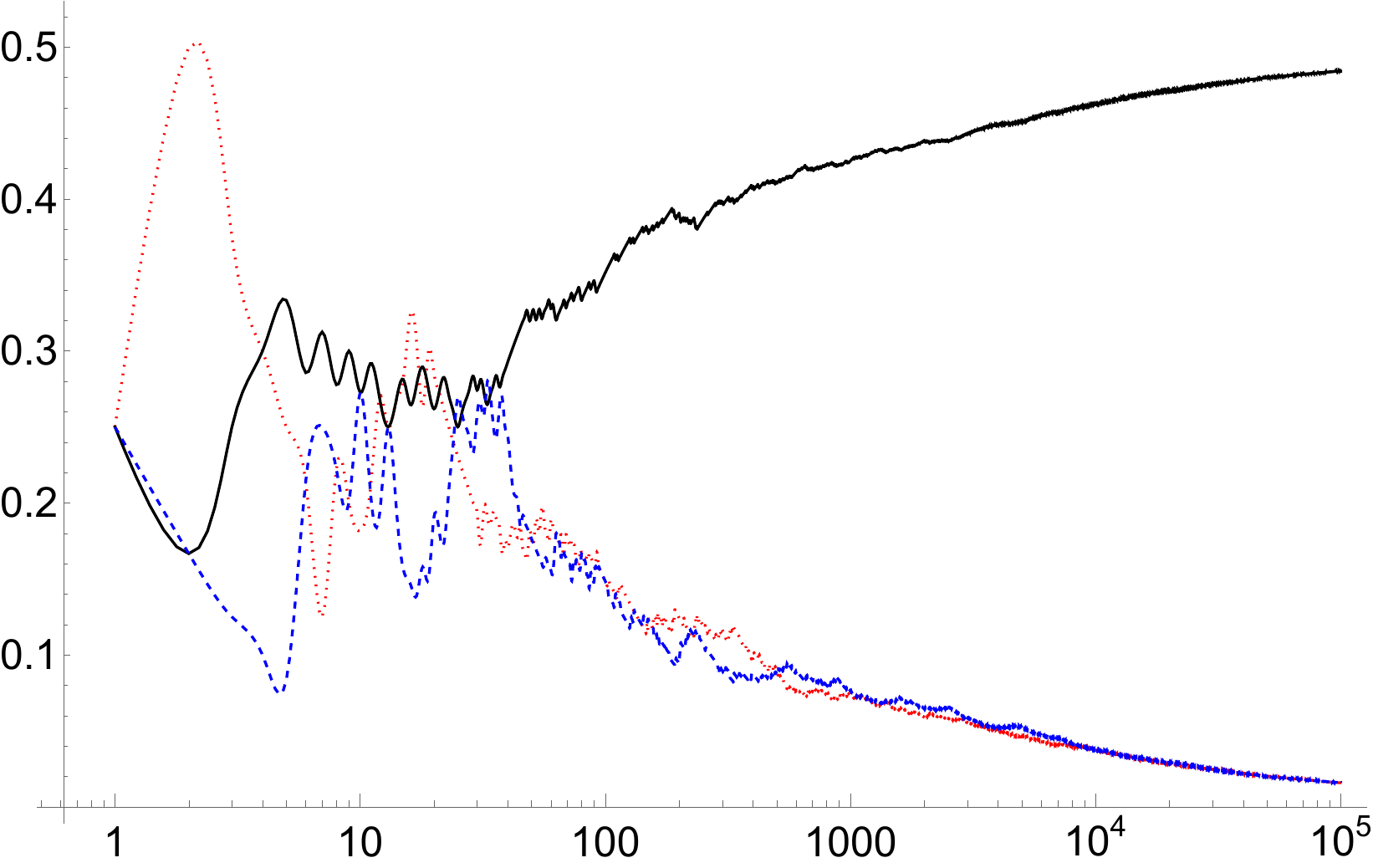}
    \put(18,55){(a)}
    \put(55,55){(b)}
    \put(22,10){(c)}
    \put(52,-2){$n$}
    \end{overpic}\\ 
    \ \\
    \caption{An example simulation of the reverse-complement string-duplication system with starting sequence $x=00$, and duplication length $k=2$, showing for $S_n(00)$ the frequencies of the substrings (a) $00$, (b) $01$ (which equals that of $10$), and (c) $11$.}
    \label{fig:rc}
\end{figure}
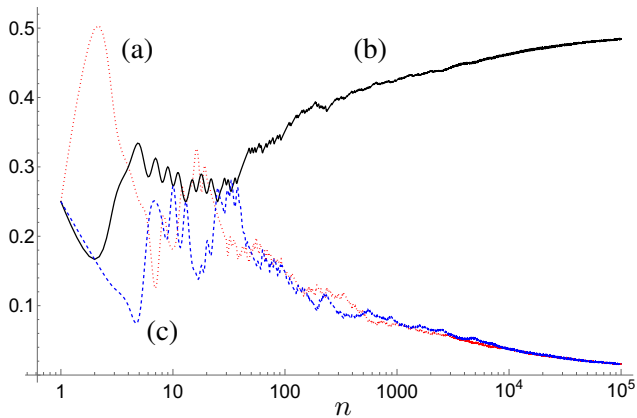

Once again we demonstrate the intricacies of string-duplication systems by showing how even the smallest of changes create dramatically different results. For the sake of this demonstration we focus on the reverse-complement string duplication system over the binary alphabet $\Sigma=\set{0,1}$. We further assume for simplicity that the initial string is $x=0$, all duplications are of the same fixed length $k=1$, and that their location is chosen uniformly and independently in each round. We emphasize that the fact the locations are chosen uniformly does not mean that $S_n(x)$ is distributed uniformly. For example, there is only one way of deriving $0111$ from $x=0$,
\[ 0 \der \underline{0}\overline{1} \der \underline{0}\overline{1}1 \der \underline{0}\overline{1} 11,\]
and the probability of this happening is exactly $1\cdot \frac{1}{2}\cdot\frac{1}{3} = \frac{1}{6}$. However, there are two ways of deriving $0101$
\begin{align*}
0 & \der \underline{0}\overline{1} \der \underline{0}\overline{1}1 \der 0\underline{1}\overline{0}1, \\
0 & \der \underline{0}\overline{1} \der 0\underline{1}\overline{0} \der 01\underline{0}\overline{1},
\end{align*}
and we get $0101$ with probability $\frac{1}{3}$. Interestingly, the entropy-rate we are after is connected to the asymptotics of permutation signatures, and the best we know  is that for duplication length $k=1$,
\[ 0.8689 \leq h(S(x=0)) \leq 0.9067 < \ccap(x=0)=1.\]
Importantly, while the capacity is full (i.e., ``most'' sequences are obtainable via carefully chosen derivation paths), not all outcomes are probable, and the entropy rate is strictly less than $1$. To further complicate matters, consider duplication of length $k=2$ and an equally long starting sequence, $x=00$. In this case, it has been shown that
\[ 0 = h(S(x=00)) < \ccap(x=00)=1.\]
Namely, while again, ``most'' sequences are obtainable, only very few are probable. This surprising result was obtained by proving that with high probability, $S_n(x=00)$ is eventually almost entirely an alternating sequence of $0101\dots$. A simulation of this fact is shown Figure~\ref{fig:rc}.

The last two properties we shall describe are perhaps more interesting from a bio-informatics perspective. Consider the following question: We are given some distant ancestor that humans evolved from, and this ancestor does not have the DNA substring that codes for a specific protein humans have. Can tandem duplication \emph{alone} mutate the ancestor's DNA sequence into a sequence that contains the instructions for that protein? In our mathematical framework of string-duplication systems, we say a system with a starting sequence $x\in\Sigma^*$ is \emph{fully expressive} if any given sequence $y\in\Sigma^*$ appears as a substring of some descendant of $x$. Returning to our previous example of end duplication vs.~tandem duplication (both of some fixed length $k$), it was shown that tandem duplication is not fully expressive, whereas end duplication is. To put this in context, imagine the following challenge: We are given Tolstoy's ``War and Peace'' (which we consider as a very long sequence of symbols). We can duplicate substrings of some fixed length, say, $k=200$. Our goal is to create a substring which is Shakespeare's ``Macbeth''. When the duplicated parts are inserted next to their original position (tandem duplication), this is impossible. However, when the duplicated parts are placed at the end, the challenge is solvable! (albeit, the procedure might be extremely lengthy)

The final property we would like to mention is that of distance to the root. When reading a mutated version $y\in\Sigma^*$ of the stored sequence $x\in\Sigma^*$, our goal is to reverse the mutation process and find $x$. This may be performed by undoing duplications, a process called \emph{de-duplication}. The process stops when no further de-duplications are possible, and the sequence at that point is called a \emph{root} of $y$. The number of de-duplication steps is called the distance to a root from $y$. In the binary case $\Sigma=\set{0,1}$, with unbounded tandem duplication, the root is always one of six options $\set{0,1,01,10,010,101}$. In this setting, we let $f(n)$ denote the maximum distance to the root of a binary sequence of length $n$. Surprisingly,
\[ 0.045 \leq \lim_{n\to\infty}\frac{f(n)}{n} \leq 0.4,\]
and the lower bound in fact holds for all but an exponentially small fraction of sequences of length $n$. Thus, the vast majority of sequences have a distance to the root that is linear in their length. This result remains essentially the same even if the duplication process is imprecise.

\subsection{Constructing error-correcting codes}

Armed with a better understanding of descendant cones, we may now approach the problem of designing error-correcting codes for string-duplication channels. Two main issues are of interest: finding a good code (i.e., making sure descendant cones of distinct codewords are disjoint), and finding an efficient decoding algorithm. Throughout this section, we shall consider the tandem-duplication string-duplication channel as an example.

When the duplication length is fixed at some length $k$, we are indeed fortunate. If the alphabet contains exactly $q$ letters, we may assume without loss of generality that $\Sigma=\Z_q$, namely, the ring of integers with addition modulo $q$. It has been suggested to view any string after taking the $k$-step discrete derivative $\partial_k$, i.e., for each $i$, subtracting the letter in position $i-k$ from the letter in position $i$. Thus, $\partial_k x = x0^k-0^kx$, where $0^k$ denotes a run of $k$ zeros, and subtraction is symbol-wise over $\Z_q$. This operation is invertible. Moreover, in the derivative domain, a tandem duplication of length $k$ manifests as an insertion of $0^k$. As a consequence, we obtain the following:
\begin{itemize}
    \item Any sequence $x\in\Sigma^*$ has a unique root.
    \item
    The unique root of a sequence $x$ may be reached by de-duplications performed in any order.
    \item
    Two sequences, $x$ and $x'$, have intersecting descendant cones if and only if they have the same root.
\end{itemize}
We note in passing that these assertions are not true even if we relax our setting minutely. For example, if we allow de-duplications of any length (instead of a fixed length) then the sequence $210121010$ does not have a unique root:
\begin{align*}
    2101\underline{2101}0 & \dedup 210\underline{10} \dedup 210, \\
    2101210\underline{10} & \dedup 2101210,
\end{align*}
where $\dedup$ denotes a de-duplication, and the underlined part is de-deuplicated.

Returning to our search for codes that correct tandem duplications of fixed length $k$, the three properties listed above lead us to the following solution: Construct a code by taking as codewords all the irreducible sequences of length $n$ over $\Z_q$ (where an irreducible sequence is a sequence that is its own root, namely, it does not contain any duplications of length $k$). With a small tweak this code can be made optimal, and allows information storage at a rate of
\[ \log_2 q - \frac{(q-1)\log_2 e}{q^{k+1}}(1+o(1)).\]
Decoding is simple, since by the properties above, we can de-duplicate in any order we wish, until reaching the unique root which must be the transmitted sequence.

In essence, since the duplications introduced by biological processes are part of the evolutionary process, we can think of the error-correcting codes we described as evolution-correcting codes.

\section{Bibliographic Notes}
There are several implementations \cite{antkowiak2020low,bornholt2016dna,blawat2016forward,chandak2019improved,church2012next,davis1996microvenus,erlich2017dna,grass2015robust,goldman2013towards,organick2017scaling,ShiNivMacChu17,tabatabaei2015rewritable,wong2003organic,yazdi2017portable,yim2021robust} of in-vitro and in-vivo DNA storage demonstrating their potential and motivating the error and channel models considered in this paper. A detailed description of the errors and the channel can be found in \cite{heckel2019characterization,Lanetal01}. 
We also refer to \cite{dong2020dna,shomorony2022information,yazdi2015dna} for a broader overview of different aspects in DNA storage.
\subsection{Deletion Codes}
The study of codes correcting deletions and insertions was introduced in the seminal papers \cite{sellers1962bit,levenshtein1966binary}, where it was shown in \cite{levenshtein1966binary} that deletion codes correct a combination of deletions and insertions. The upper and lower bounds on the optimal redundancy of deletion codes were also given in \cite{levenshtein1966binary}.
The VT codes were proposed in \cite{vt1965}. An algebraic generalization of VT codes with redundancy linear in code length was presented in \cite{helberg2002multiple} and further extended in \cite{hagiwara2016ordered}. The first codes correcting a number of linear in code length deletions based on concatenated code structures were proposed in \cite{schulman1999asymptotically} and were improved in \cite{guruswami2017deletion,guruswami2016efficiently}. Using the concatenated code construction, \cite{brakensiek2017efficient} proposed the first codes correcting a small number of deletions with redundancy logarithmic in the code length. For two deletions, the result in \cite{brakensiek2017efficient} was improved by \cite{gabrys2017codes,sima2019two,guruswami2021explicit}. The first order-wise optimal codes correcting a constant number of deletions were given by \cite{cheng2022deterministic,sima2020optimal,sima2020systematic}.

The algebraic generalizations of VT codes discussed in Section \ref{sec:delcodes} for correcting deletions were  presented in \cite{sima2020optimal,sima2020systematic,sima2019two}. Compared to the VT generalizations in \cite{hagiwara2016ordered,helberg2002multiple} that require linear redundancy, the generalizations in \cite{sima2020optimal,sima2020systematic,sima2019two} are capable of correcting a constant number of deletions with asymptotically at most $4$ times the optimal redundancy, which is a logarithm of the code length. Combining the codes in \cite{sima2020systematic} and the VT codes, \cite{song2021multiple} further improved the redundancy in \cite{sima2020systematic} from $4t\log n+o(\log n)$ to $(4t-1)\log n +o(\log n)$, where $t$ is the number of deletions and $n$ is the code length. Besides the above code constructions, 
existential bounds improving the results in \cite{levenshtein1966binary} were recently presented in \cite{alon2022logarithmically}.

Other related problems include: systematic deletion codes \cite{belazzougui2015efficient,cheng2022deterministic,cormode1999communication,haeupler2019optimal,orlitsky1993interactive,sima2020systematic}, non-binary deletion codes \cite{cullina2014improvement,haeupler2017synchronization,liu2022bounds,sima2020optimal,tenengolts1984nonbinary,levenshtein2002bounds,le2015new}, deletion codes with randomized decoding \cite{belazzougui2016edit,chakraborty2015low,irmak2005improved,jowhari2012efficient}, channel capacity of deletion channels \cite{cheraghchi2019capacity,diggavi2001transmission,dobrushin1967shannon,drinea2007improved,tal2021polar,venkataramanan2013achievable,kalai2010tight,kanoria2013optimal,kirsch2009directly,ullman1967capabilities}, codes correcting a combination of deletions, insertions, substitutions, and transpositions \cite{cai2021correcting,gabrys2017codes,gabrys2022beyond,cheng2018block}, codes correcting a burst of deletions \cite{wang2022permutation,schoeny2017codes,lenz2020optimal,sun2023improved}, codes for sticky insertions \cite{dolecek2010repetition,mahdavifar2017asymptotically}, and codes correcting asymmetric deletions \cite{tallini2022deletions,wang2023codes}. 
 In addition, the application of edit distance in natural language processing and biological data analysis can be found in \cite{sankoff1983time,editdistance}. 
See \cite{cheraghchi2020overview,mitzenmacher2009survey,mercier2010survey,sloane2002single} for a broader review of this topic.
\subsection{Sliced Channel}
%The workflow of DNA storage systems were implemented in experiments \cite{organick2017scaling,yazdi2017portable}. 
The model of encoding information into a set of unordered and equal-length sequences was introduced in \cite{heckel2017fundamental}, where it was shown that index-based schemes achieve the channel capacity when there are sequence losses. Later, the channel capacity analysis was extended to channels with both sequence loss and substitution errors  \cite{lenz2019upper,lenz2020achieving,shomorony2021dna,weinberger2021dna}. The protection of indices against errors assuming index-based schemes was addressed in \cite{lenz2019anchor,shinkar2021clustering,song2020sequence}. 

The definition of redundancy measuring the extra redundancy needed for error protection was introduced in \cite{lenz2019coding}, where it was shown that the redundancy for index-based schemes is linear in the number of sequences. The order-wise optimal redundancy under this definition was obtained in \cite{sima2021coding}, where the idea of using data for indexing was proposed. 
By using this idea, code constructions were presented in \cite{sima2020robust} to achieve order-wise optimal redundancy and in \cite{wei2021improved} to obtain improved results upon those in \cite{lenz2019coding}. 

Other related problems include: permutation channels \cite{kovavcevic2015perfect,kovavcevic2018codes,langberg2017coding,makur2020coding,tang2023capacity,walsh2008capacity}, codes for reconstruction from substrings \cite{chang2017rates,gabrys2018unique,kiah2016codes,raviv2018rank,yehezkeally2022generalized}, torn-paper channels \cite{bar2022adversarial,ravi2021capacity,shomorony2021torn}.
\subsection{Duplication}
Early attemps of in-vivo information storage can be found in  \cite{davis1996microvenus,wong2003organic}. 
Proofs of concepts for storing information in the DNA of living organisms were provided in~\cite{ShiNivMacChu17} and \cite{yim2021robust}. In~\cite{Lanetal01}, it was pointed out that the majority of the human DNA contains duplications. The string-duplication channels were introduced in~\cite{FarSchBru16} and the capacity and expressiveness of several duplication rules were studied. These results were extended in~\cite{JaiFarBru17,Kov19}. The stochastic channel model, called a P{\'o}lya string model, was introduced in~\cite{EliFarSchBru19}, and further studied in~\cite{FarSchBru19,LouSchBruFar20,BenSch22}. In particular,~\cite{FarSchBru19} developed a parameter-estimation scheme based on this model. The distance to the root in tandem duplication channels was studied in~\cite{AloBruFarJai17}.

Error-correcting codes for string-duplication channels were first studied in~\cite{JaiFarSchBru17a}. The work was followed by many others, among them works studying: tandem duplication \cite{KovTan18,LenWacYaa19,ZerEsmGul19,CheChrKiaNgu20,ZerEsmGul20}, Levenshtein reconstruction for uniform tandem duplication~\cite{YehSch20,YehSch21}, noisy tandem duplication~\cite{TanYehSchFar20,TanFar21,TanFar21a,TanWanLouGabFar23}, palindromic duplications~\cite{LenWacYaa19,ZerEsmGul20}, and reverse-complement duplications~\cite{BenSch22}.

\section{Acknowledgments}
This research was partially supported by NSF grant CCF-1816965 and NSF grant CCF-1717884.

\bibliographystyle{IEEEtranS}
\bibliography{allbib,bib2,bib3}
\end{document}